# Experimental verification of band convergence in Sr and Na codoped PbTe

Yuya Hattori[1, 2, a)], Shunsuke Yoshizawa[3, b)], Keisuke Sagisaka[3], Yuki Tokumoto[2], Keiichi Edagawa[2], Takako Konoike[1], Shinya Uji[1] and Taichi Terashima[1]

[1]*Research Center for Materials Nanoarchitectonics (MANA), National Institute for Materials Science, 3-13 Sakura, Tsukuba, Ibaraki 305-0003, Japan*

[2]*Institute of Industrial Science, The University of Tokyo, Komaba, Meguro-ku, Tokyo 153-8505, Japan*

[3]*Center for Basic Research on Materials, National Institute for Materials Science, 1-2-1 Sengen, Tsukuba, Ibaraki 305-0047, Japan.*

Corresponding authors' E-mail: (a): HATTORI.Yuya@nims.go.jp (Yuya Hattori), (b): YOSHIZAWA.Shunsuke@nims.go.jp (Shunsuke Yoshizawa)

**Abstract**

Scanning tunneling microscopy and transport measurements have been performed to investigate the electronic structure and its temperature dependence in heavily Sr and Na codoped PbTe, which is recognized as one of the most promising thermoelectric materials. Our main findings are as follows: (i) Below $T = 4.5$ K, all carriers are distributed in the first valence band at the $L$ point ($L$ band), which forms tube-shaped Fermi surfaces with concave curvature. With Sr and Na doping, the dispersion of the $L$ band changes, and the band gap increases from $E_G \cong 200$ meV to 300 meV. (ii) At $T = 4.5$ K, the Fermi energy is located $\sim 100$ meV below the edge of the $L$ band for the Sr/Na codoped PbTe. The second valence band at the $\Sigma$ point ($\Sigma$ band) is lower than the $L$ band by $\Delta E_{L\Sigma} \cong 150$ meV, which is significantly smaller than that of pristine PbTe ($\Delta E_{L\Sigma} \cong 200$ meV). The decrease in $\Delta E_{L\Sigma}$, leading to band convergence, provides a desirable condition for thermoelectric materials. (iii) With increasing temperature, the carrier distribution to the $\Sigma$ band starts at $T \cong 100$ K, and we estimate that about 50% of the total carriers are redistributed in the $\Sigma$ band at $T = 300$ K. Our work demonstrates that scanning tunneling microscopy and angular dependent magnetoresistance measurements are particularly powerful tools to determine the electronic structure and carrier distribution. We believe that they will provide a bird's eye view of the doping strategy towards realizing high-efficiency thermoelectric materials.

# I. INTRODUCTION

Since the 2010s, the performance of thermoelectric (TE) materials has been improving drastically [1] [2]. The conversion efficiency of the thermoelectricity is given by the thermoelectric figure of merit $ZT = \sigma S^2 T/\kappa$, where $\sigma$, $S$, $T$, and $\kappa$ are the electrical conductivity, Seebeck coefficient, temperature, and the thermal conductivity, respectively. Until the 2000s, the $ZT$ values of the available bulk TE materials were mostly less than $ZT = 1$; however, many materials with $ZT > 2$ have been discovered since the 2010s [3] [4] [5] [6]. One of the most promising and widely used strategies for synthesizing high-performance TE materials is band convergence [7] [8] [9]. By the Mott expression, the Seebeck coefficient $S$ is calculated as follows [10] [11] [12]:

$$S = -\frac{\pi^2 k_B^2 T}{3|e|}\left\{\frac{1}{N}\frac{dN(E)}{dE} + \frac{1}{\mu}\frac{d\mu(E)}{dE}\right\}_{E=E_F}. \qquad (1)$$

Here, $N(E)$, $\mu(E)$, and $E_F$ are the density of states (DOS), mobility, and the Fermi energy respectively. This formula tells us that materials with high $dN(E)/dE$ values at $E = E_F$ would have a high Seebeck coefficient. This condition is realized in band-converged systems, in which the Fermi energy is located near the band edge of the second valence/conduction band, resulting in a large increase in $dN(E)/dE$. In Sr/Na codoped PbTe, for example, both the first valence band at the $L$ point with light mass and low valley degeneracy ($L$ band, $N_L = 4$), and the second valence band at the $\Sigma$ point with heavy mass and high valley degeneracy ($\Sigma$ band, $N_\Sigma = 12$) are located near the Fermi energy [3] [8]. The high electrical conductivity is attributed to the $L$ band with light mass, and the large Seebeck coefficient originates from the $\Sigma$ band with high valley degeneracy. The figure of merit reaches as high as $ZT = 2.5$ in this system [8]. The band convergence strategy has also been applied to many other systems such as SnTe [13] [14],

CoSb$_3$ [15], and ZrCoBi [16], and their TE performances have been greatly improved.

Although band convergence is a good practical strategy for improving the TE efficiency, the reason why it leads to high values of $ZT$ still remains unclear [17] [18]. While the Seebeck coefficient $S$ is enhanced in band-converged systems according to Eq. (1), the mobility $\mu$ tends to decrease in some materials owing to the increased scattering rate between multiple bands [17], resulting in low conductivity $\sigma$ [18]. In addition, the thermal conductivity of electrons $\kappa_{el}$ should increase by band convergence due to the increased DOS. Therefore, it is not self-evident why band convergence is an effective strategy for realizing the high TE performance.

To discuss the correlation between the TE performance and electronic structure and to apply it to the synthesis strategy of TE materials, it is crucial to determine the band structure and carrier distribution in doped systems and also at high temperatures. However, it is difficult to elucidate them both theoretically and experimentally. In first-principles calculations, the band gap and band offset, which are important factors for the TE performance, are sensitive to calculation methods and input parameters even in pristine systems [19] [20]. Furthermore, these values are known to be strongly modified with doping [9] [21] and temperature [22] [23]. Although extensive studies based on transport measurements have been performed in polycrystalline and doped samples of TE materials at high temperatures, the experimental determination of the band structure and carrier distribution in TE materials is a surprisingly difficult task. Therefore, other techniques for the experimental determination of electronic structure and carrier distribution are urgently needed.

In this study, we combined scanning tunneling microscopy (STM) and transport measurements to investigate the electronic structure and its temperature dependence in Sr/Na codoped PbTe. Our work demonstrates that DOS measurements by STM and the analyses of the angular dependent magnetoresistance (AMR) are particularly powerful tools to determine the electronic structure of TE materials experimentally, and we believe that they will provide a bird's-eye view of the doping strategy of TE materials.

## II. METHOD

We synthesized single crystals of Sr/Na codoped PbTe and pristine PbTe by the vertical Bridgman method. In the preparation process of Sr/Na codoped PbTe, a mixture of Pb (purity: 99.9999 %), Te (99.9999 %), Sr (99%), and Na (99.95%) was sealed in a quartz tube with a molar ratio of Pb:Te:Sr:Na = 96:100:4:2. The mixture was melted in a furnace at $T = 1323$ K and air-cooled at $T = 293$ K to obtain polycrystalline samples. The Bridgman method was then applied to some of the ingots to obtain single crystalline samples. Single crystalline samples of pristine PbTe were synthesized from a mixture of Pb and Te with a molar ratio of Pb:Te = 45:55 to decrease the melting point and stabilize the crystal growth. The remaining procedures were the same with that of Sr/Na codoped PbTe. The crystal structure and composition were investigated by powder x-ray diffraction (XRD; RIGAKU RINT2500V) and electron probe micro analyzer (EPMA; JEOL JXA-8800RL).

The density functional theory (DFT) calculation of the band structure of pristine PbTe was carried out using the OpenMX code [24] within the Perdew-Burke-Ernzerhof generalized gradient approximation [25]. The basis functions were pseudo atomic orbitals (PAO) provided as Pb10.0-s3p3d2f1 and Te7.0-s3p3d2f1. The energy cutoff was 300 Ry, the $k$-point mesh was 16×16×16, and the spin-orbit coupling was included. The experimental value of 6.46 Å [26] was adopted for the lattice constant of the conventional unit cell (Pb-Te-Pb distance), and the calculations were performed using the primitive cell.

STM measurements were performed using an ultra-high vacuum cryogenic STM (Unisoku USM-1300). Single-crystalline samples were cleaved by pushing a blade in an ultrahigh-vacuum environment to expose clean surfaces and transferred immediately to the STM head kept at a low temperature. The STM measurements were performed at $T = 4.5$ K. Differential conductance ($dI/dV$) curves were obtained by using a lock-in technique with a modulation voltage of 5 mV at 831 Hz for Sr/Na codoped PbTe and by numerical differentiation of the $I-V$ curves for pristine PbTe.

Resistivity and Hall measurements were performed by standard four-contact resistance measurements from $T = 1.4$ K to 300 K in magnetic fields up to $B = 14.5$ T. The electrical contacts were achieved by spot welding. The direction of the current $I$ is along the [100] direction in all the measurements in this study. In the resistivity measurements, positive and negative magnetic fields were applied, and the raw data were numerically symmetrized/antisymmetrized to purely extract magneto resistance (MR)/Hall resistivity. The AMR measurements were carried out by measuring the angular dependence of the electrical resistance at $B = \pm 14.5$ T and the data were averaged to eliminate the component of the Hall resistivity.

### III. RESULTS AND DISCUSSION
#### A. Sample characterization

The powder XRD analyses of the single crystals of Sr/Na codoped PbTe and pristine PbTe show that all the peaks can be indexed with the structure of PbTe and that the samples do not contain any impurity phase [see Supplementary Material (SM) [27]]. Table I lists the EPMA results of a single crystal of Sr/Na codoped PbTe (Bridgman method) together with that of polycrystalline Sr/Na codoped PbTe as a reference. The chemical composition obtained by EPMA is an average of five different probe spots. The polycrystalline sample was obtained by air-cooling, which is a typical pretreatment of PbTe in studies on TE properties [3]. As shown in Table I, the concentrations of Sr and Na in our single crystal are comparable to those in the bulk matrix of polycrystalline samples. This agreement ensures that the results obtained in this study can be applied to the study of PbTe in the context of TE materials. We note that dendritic microstructure with high Sr concentration and small spots with high Na concentration were found in the air-cooled samples (see SM [27]). This microstructure explains the lower Sr and Na concentrations than the nominal composition.

### B. Band calculation

The calculated band structure of pristine PbTe is shown in Fig. 1(a). There is a direct band gap at the $L$ point, and the edge of the second valence band is located along the $\Gamma - K$ direction, as reported in earlier studies [30] [41]. Figure 1(b) shows the calculated Fermi surface (depicted using FermiSurfer software [42]) with the Fermi energy adjusted manually to $E_F = 110$ meV below the $L$ band edge. There are two types of the Fermi surfaces: a tube-shaped hole pocket at the $L$ point and banana-shaped hole pocket at the $\Sigma$ point. As shown in the earlier studies [31] [33], the $L$ band dominates the total electrical conduction at low temperatures. In samples with high hole concentrations and at high temperatures, the $\Sigma$ band is considered to dominate the electrical conduction as it approaches the Fermi energy at high temperatures [31] [32]. In the earlier studies, the carrier redistribution to the $\Sigma$ band was inferred from the increase of the effective mass at high temperatures [43] and the temperature dependence of the Hall coefficient [31] [38]. However, the position of the Fermi energy $E_F$ and detailed band structure of doped PbTe are still controversial as we will discuss in Sec. III D.

### C. STM measurements

In Fig. 2(a), the STM topography on the (001) surface of single-crystalline Sr/Na codoped PbTe is presented. A clean surface with atomic resolution was successfully obtained by cleaving. The topography shows some bright spots of impurity sites as well as void defects of vacancy sites. The former are probably the dopants precipitated on the surface, while the latter could be generated during the cleavage of the sample. Figure 2(b) shows the $dI/dV$ curves plotted on a semi-logarithmic scale. The measurement positions are indicated in Fig. 2(a) by markers in the same colors as the $dI/dV$ curves. The $dI/dV$ curves correspond to the local DOS on the sample surface and provide information about the band structure and band gap [44] [45] [46]. In Fig. 2(b), we notice remarkably low $dI/dV$ values between $V \cong +100$ mV and $+400$ mV, which is ascribed to the band gap of this material, $E_G \cong 300$ meV. This band profile shows p-type conduction (hole carriers) with $E_F \cong 100$ meV from the edge of the $L$ band. Further, in the figure, we also find that the $dI/dV$ curves have a kink at $V \cong -50$ mV. This kink is attributed to the edge of the $\Sigma$ band. Then, the band offset $\Delta E_{L\Sigma}$ is determined as $\Delta E_{L\Sigma} \cong 150$ meV, which is defined as the energy difference between the band edge at the $L$ point and that of the $\Sigma$ point as indicated in Fig. 1(a).

Figure 2(c) shows the $dI/dV$ spectrum measured on the (001) surface of single-crystalline pristine PbTe at $T = 4.5$ K. Similarly, we observe the band gap $E_G \cong 200$ meV from $V \cong -200$ mV to 0 mV, which is in good agreement with the optical absorption measurements in the earlier studies [40] [47] ($E_G \cong 200$ meV). We obtain the larger band offset $\Delta E_{L\Sigma} \cong 200$ meV from the kink at $V \cong -400$ mV. Note that the $dI/dV$ spectrum of pristine PbTe obtained here (the linear plot in the wider bias range is shown in SM [27]) agrees well with the earlier theoretical study [30], although there is no structure from resonances in our $dI/dV$ spectrum.

In the previous studies, the high TE performance of Sr/Na codoped PbTe has been attributed to band convergence [3] [8], but this explanation is based only on band calculations. The $dI/dV$ spectra obtained in this study clearly show that the band offset $\Delta E_{L\Sigma}$ in Sr/Na codoped PbTe becomes certainly smaller than that in pristine PbTe, thereby providing a more favorable condition for band convergence [7] [14]. The smaller band offset $\Delta E_{L\Sigma}$ is also qualitatively consistent with the temperature dependence of the Hall coefficient reported in Sr/Na codoped PbTe polycrystalline samples [8]. In addition, notably, the band gap increases significantly from $E_G \cong 200$ meV to $E_G \cong 300$ meV with Sr and Na doping. The energy separation between the Fermi energy and the bottom of the conduction band becomes roughly 400 meV in the Sr/Na codoped samples. Because of the large energy separation, the necessary condition to minimize the reduction of the Seebeck coefficient $S$ by bipolar conduction, i.e., $E_G + E_F > 5k_BT$ [30] [48], is satisfied even at $T = 900$ K ($5k_BT \cong 388$ meV) in Sr/Na codoped PbTe. This condition is advantageous for achieving high ZT values at high temperatures.

One of the advantages of $dI/dV$ measurements by STM is the ability to obtain information about the unoccupied parts of the band structure. In their pioneering work, Takeuchi et al. [49] revealed the relationship between the TE performance and electronic structure in layered cobalt oxides by photoemission spectroscopy (PES) measurements. Note that the information of the unoccupied bands, which can be deduced from the normalized data by the Fermi distribution function, is not so reliable in PES due to the low signals of the unoccupied bands. Since the TE properties are generally determined by the electronic states in the range of $\zeta - 5k_BT < E < \zeta + 5k_BT$ [30] [48], where $\zeta$ is the chemical potential and $\zeta \cong E_F$ at low temperatures, the unoccupied part of the band structure is as important as the occupied part. Therefore, we believe that the $dI/dV$ measurement by STM is an ideal tool to discuss the correlation between the band structure and TE performance.

The Fermi energy determined by our $dI/dV$ measurements agrees well with the earlier study of heavily Na-doped PbTe with high hole concentrations [31]; this study showed that the total carriers are located only in the $L$ band at $T = 4.2$ K. However, in a recent study on PbTe with the Na solubility limit, specific heat measurements indicated that the Fermi energy at $T \cong 1$ K also crosses the $\Sigma$ band [50]. The above inconsistency requires a detailed study of the Fermi energy and Fermi surface, such as quantum oscillation measurements as shown in Sec. III D.

### D. Magnetoresistance and quantum oscillation

Figure 3(a) shows the field dependence of the MR at $T = 1.4$ K measured at different field angles for single-crystalline Sr/Na codoped PbTe, where $\rho(B)$ is the electrical resistivity at the magnetic field $B$, and $\rho_0$ represents that at $B = 0$. The definition of the angle $\theta_1$ is shown in the inset. At low fields, the MR curves fall into almost the same curve and the field dependence at low magnetic fields is approximately $\rho_{xx} \propto B^{1.6}$, which is similar to the typical behavior of metals at low fields: $\rho_{xx} \propto B^2$ [51]. At high fields, they deviate from each other, and an angle-dependent oscillatory behavior is also evident. Figure 3(b) shows the oscillatory components of the MR, namely the Shubnikov-de Haas (SdH) oscillations, which are obtained by subtracting smooth backgrounds from the MR. Such large SdH oscillations for the heavily doped PbTe are surprising, but it could be ascribed to the nearly complete screening of impurity potentials by the high dielectric constant of PbTe [52]. It was previously reported that the scattering time $\tau$ is insensitive to the dopant concentration in the SdH measurements of lightly Na-doped PbTe [53].

Figures 3(c) and 3(d) show the Fourier spectra of the SdH oscillations in magnetic field from [001] to [010] (angle step: 5°), and from [001] to [110] (angle step: 2°), respectively. The definitions of the field angles $\theta_1$ and $\theta_2$ are given in the insets of Figs. 3(c) and 3(d), respectively. Previous studies [39] [50] showed that the Hall coefficient $R_H$ at low fields in PbTe with high hole concentrations is not directly related to the carrier concentration $n$. In this case, the frequency of the SdH oscillation is a good indicator of the carrier concentration. In our sample, the frequency of the SdH oscillations is $F_{111} = F_{min} = 104$ T for $\boldsymbol{B}//[111]$ and $F_{001} = 180$ T for $\boldsymbol{B}//[001]$. These values are slightly larger than those for Na-doped PbTe [53] ($F_{111} = 97$ T, $F_{001} = 157.5$ T, $n = 8.3 \times 10^{19}$ cm$^{-3}$). Then, the total carrier concentration of the Sr/Na codoped PbTe is roughly expected to be $n \cong 1.0 \times 10^{20}$ cm$^{-3}$ considering the relationship between the carrier concentration $n$ and the SdH frequency $F$—that is, $n \propto F^{1.5}$. We will later discuss the relationship between the carrier concentration $n$ and Hall coefficient $R_H$ in Sec. III E.

As can be seen in Fig. 3(c), the frequency $F$ takes almost the same value at $\theta_1 = 0°$ and $\theta_1 = 90°$, which is consistent with the cubic symmetry of the crystal (space group: 225, $Fm\bar{3}m$). In PbTe, the Fermi surface at the $L$ point has been assumed to have an ellipsoidal shape with its longer axis along the [111] direction [33] [54]. For the field rotation from [001] to [110] in Fig. 3(d), the SdH frequency $F$ takes the minimum value when $\boldsymbol{B}//[111]$ ($\theta_2 \sim 54°$), which is consistent with the ellipsoidal Fermi surface at the $L$ point [54]. In Figs. 3(c) and 3(d), we find no trace of the $\Sigma$ band, which should have the extremum of $F$ at $\theta_1 = \theta_2 = 0°, 90°$ [53]. These results are consistent with our STM measurements shown in Fig. 2(b), which indicate that the $\Sigma$ band edge is lower than $E_F$. Our results here ensure that no carriers are distributed to the $\Sigma$ band below $T = 4.5$ K.

In Fig. 3(e), the angular dependence of the experimental SdH frequency $F$ (red crosses) is plotted together with the simulated values (blue solid curves). The simulations were performed assuming an ellipsoidal Fermi surface at the $L$ point with an anisotropy parameter of $K = (c/a)^2 = 15$ [53], where $a$ and $c$ are the shorter and longer axes of the ellipsoid, respectively. The agreement between the experimental and the simulated $F$ seems satisfactory, but a slight deviation is observed at the field directions far from the [111] direction ($\theta_2 \cong 54.7°$), as shown in the inset of Fig. 3(e).

In order to evaluate the shape of the Fermi surface more precisely and quantitatively, we follow the analysis of Walmsley et al. [50]. If the Fermi surface is a perfect ellipsoid with an anisotropy parameter $K_0$, $F$ can be calculated as follows:

$$F(\varphi, K_0) = F_{min} K_0^{0.5} (K_0 \cos^2 \varphi + \sin^2 \varphi)^{-0.5}, \tag{2}$$

where $F_{min} = F_{111}$ is the minimum frequency and $\varphi$ is the angle from the [111] direction. Then, the shape function $G(\varphi, K_0)$ below would be unity at any $\varphi$ if the Fermi surface is perfectly ellipsoidal with the anisotropy parameter $K = K_0$.

$$G(\varphi, K_0) = \frac{F_{exp}}{F(\varphi, K_0)} = \frac{F_{exp}}{F_{min} K_0^{0.5}} (K_0 \cos^2 \varphi + \sin^2 \varphi)^{0.5}, \tag{3}$$

where $F_{exp}$ is the experimental value of the SdH frequency. We set $F_{min} = 104$ T based on our experimental data, and $K_0 = 15$ following the previous study of SdH measurements [53]. Then, $G(\varphi, K_0)$ is plotted as shown in Fig. 3(f). If the Fermi surface is an ellipsoid with $K > 15$, $G(\varphi, K_0)$ is greater than unity, but it is less than unity for that with $K < 15$ as shown in Fig. 3(f). The plot of the experimental data in Fig. 3(f) explicitly shows that the Fermi surface of Sr/Na codoped PbTe is not an ellipsoidal Fermi surface as observed in the SdH measurements of lightly Na-doped PbTe [53]. Rather, the Fermi surface is likely to be cylindrical with a slight dent in the center. This shape is similar to that reported in heavily Na-doped PbTe [50]. As we discuss in Sec. III E, the concave curvature of the Fermi surface strongly affects the Hall resistivity $R_H$ and Seebeck coefficient $S$.

From the temperature dependence of the SdH oscillations (see SM [27]), we obtained the effective mass $m_c^* = 0.11 m_e$ ($m_e$: the free electron mass) along the [111] direction and $m_c^* = 0.21 m_e$ along the [100] direction in the Sr/Na codoped PbTe. The effective mass is comparable to the values reported for Na-doped PbTe [53] ($m_{111}^* = 0.13\ m_e$, $m_{100}^* = 0.23\ m_e$), and the mobility $\mu$ of the $L$ band apparently remains high upon Sr doping because the scattering time $\tau$ in PbTe is insensitive to the dopant concentration [53]. This is a desirable condition to achieve high values of $\sigma$ and $ZT$. In the Kane model, the electronic structure is expressed as a combination of parabolic and linear dispersions [50] [55]:

$$E\left(1 + \frac{E}{E_G}\right) = \frac{\hbar^2 k_\perp^2}{2m_\perp} + \frac{\hbar^2 k_\parallel^2}{2m_\parallel}, \qquad (4)$$

where $E_G$, $k_\perp$, $k_\parallel$, $m_\perp$, and $m_\parallel$ are the band gap, the magnitude of the transverse and longitudinal components of $\boldsymbol{k}$ measured from the $L$ point, the effective mass along the transverse and longitudinal directions, respectively. So far, the $L$ band edge of PbTe has been analyzed by the Kane model [56] [57], in which the parabolicity depends on the value of $E_G$. In Sr/Na codoped PbTe, we obtained $k_F = 0.056$ Å$^{-1}$ and $m_c^* = 0.11 m_e$ along the [111] direction from the SdH measurements. Then, we obtain $E_F =$ $\hbar^2 k_F^2 / 2m_c^* = 108$ meV below the edge of the $L$ band for the parabolic band and $E_F = \hbar^2 k_F^2 / m_c^* = 216$ meV for the linear band. The Fermi energy obtained from STM measurements ($E_F \cong 100$ meV) is much closer to the value of the parabolic band.

The parabolicity of the $L$ valence band can also be inferred from the $dI/dV$ curves in Figs. 2(b) and 2(c). As shown in the inset of Fig. 2(c), the $dI/dV$ curves below -200 meV and above 0 meV are concave upward functions in pristine PbTe. This suggests that both the valence and conduction bands at the $L$ point have a linear dispersion, because the DOS is proportional to $E^2$ when $E \propto k$. The $dI/dV$ curve below -400 meV is a concave-downward function (see SM [27]). This implies that the $\Sigma$ valence band is parabolic, because the DOS is proportional to $E^{0.5}$ when $E \propto k^2$. These results are in good agreement with an earlier study on PbTe, which showed the non-parabolic dispersion of the $L$ band [31]. In contrast, the $dI/dV$ curves in Sr/Na codoped PbTe near the first valence band edge do not follow the typical behavior of the linear 3D band (the inset of Fig. 2(b)), but it shows a behavior intermediate between parabolic ($DOS \propto E^{0.5}$) and linear band ($DOS \propto E^2$). This result is surprising, but we consider that the wider band gap $E_G$ ($E_G = 300$ meV at $T = 4.5$ K) in Sr/Na codoped PbTe makes the second term on the left hand of Eq. (4) negligible and also makes the band structure at the $L$ point more parabolic. We note that some studies based on the Kane model reported much higher Fermi energies than our estimate ($E_F = 180$ meV in Na-doped PbTe [50], $E_F = 190 - 210$ meV in Tl-doped PbTe [58]). We assume that their results differ from ours because they used the band gap of pristine PbTe ($E_G = 200$ meV) and assumed more linear band.

### E. Analyses of the AMR, Kohler plot, and Hall resistivity

In this section, the temperature dependence of the carrier distribution and band structure is discussed. In Fig. 4(a), the AMR of Sr/Na codoped PbTe at $B = 14.5$ T and at different temperatures is presented, where $\rho(\theta_1, B)$ represents the electrical resistivity with a magnetic field $B$ at the angle $\theta_1$. The magnetic field was tilted from [001] to [010] as shown in the inset of Fig. 4(a), and the direction of the current is $I//[100]$, then the MR is mainly from the transverse MR. In this measurement, the angular dependence of the electrical resistance was measured at $B = \pm 14.5$ T, and the data were averaged to exclude the component of the Hall resistivity. As seen from Fig. 4(a), the curve takes the minimum values at $\theta = 0°, 90°, -90°$ at $T = 1.4$ K. The small oscillatory components originate from the SdH oscillation. The symmetric behavior across $\theta_1 = 0°$ is consistent with the crystal symmetry. In SM [27], we show that the small deviation from the 4-fold symmetry ($\Delta\theta_1 = 90°$ periodicity) of the crystal originates from the contribution of the longitudinal MR ($\Delta\theta_1 = 180°$ periodicity [33]), which is reported to be much larger than the transverse MR in PbTe [34]. With increasing temperature, the anisotropy of the AMR is strongly suppressed and then the anisotropic behavior is reversed above $T = 200$ K as shown in the right panel of Fig. 4(a): the AMR has the minimum at $\theta_1 = 0°$ at low temperatures, but the maximum at $\theta_1 = 0°$ at high temperatures. Similar reversal behavior was observed in the MR of pristine PbTe at higher temperature: $T \cong 350$ K [33]. The reversal behavior of the AMR in Fig. 4(a) is strong evidence for carrier redistribution from the $L$ band to the $\Sigma$ band.

The AMR is closely related to the shape of the Fermi surface, and the AMR measurements have been traditionally performed to study the Fermi surface [59] [60] [61] [62]. If the magnetic field and the Fermi velocity are parallel, it does not affect the motion of electrons and the MR would not increase. If the Fermi surface is an anisotropic ellipsoid, the Fermi velocity of the majority of carriers is perpendicular to the longer direction of the Fermi surface. Then, the AMR takes its maximum value when $\boldsymbol{B}$ is parallel to the longer axis of the Fermi surface. In the case of PbTe, which has only the $L$ pocket at low temperatures [31], the AMR is expected to have its maximum when $\theta_1 = 45°$ (for $\boldsymbol{B}$ almost parallel to the longer axis of the $L$ pocket). As the temperature increases and the $\Sigma$ band dominates the electrical conduction [33], the AMR will have its maximum at $\theta_1 = 0°$ (for $\boldsymbol{B}//[001]$, the longer axis of the $\Sigma$ pocket). These expectations are consistent with the experimental results shown in Fig. 4(a). We assume that almost all carriers are located at the $L$ pocket below $T = 100$ K considering the AMR behavior. At higher temperatures ($T = 150, 200$ K), a part of the carriers are distributed to the $\Sigma$ pocket and the AMR loses its symmetric shape as shown in the right panel of Fig. 4(a). At $T = 300$ K, the AMR should be mainly contributed by the $\Sigma$ pocket, judging from the behavior of the AMR shown in Fig. 4(a). In order to quantitatively estimate the temperature dependence of the carrier distribution and band structure, we next conducted analyses of the MR and Hall resistivity.

Figure 4(b) shows the Kohler plot of Sr/Na codoped PbTe, where $\Delta\rho(B) = \rho(B) - \rho_0$. Here, the magnetic field $\boldsymbol{B}$ was applied along $\boldsymbol{B}//[001]$. If the Fermi surface structure and the scattering mechanism are independent of temperature, the MR $\Delta\rho(B)$ is known to follow Kohler's rule [63]:

$$\frac{\Delta\rho(B)}{\rho_0} = f\left(\frac{B}{\rho_0}\right). \tag{5}$$

Then if we plot $\Delta\rho(B)/\rho_0$ vs $B^2/\rho_0^2$ (the Kohler plot), the curves at different temperatures should lie on the same curve. As shown in Fig. 4(b), the curves at lower temperatures roughly match to the same curve, but the deviation becomes more pronounced at high temperatures. We attribute this deviation to the redistribution of the carriers from the $L$ pocket to the $\Sigma$ pocket [7]. Still, it should also be noted that the temperature dependence of the Kohler plot at $T < 100$ K is different from that at $100\text{ K} \leq T \leq 300\text{ K}$, as indicated by the arrows in Fig. 4(b).

Figure 4(c) shows the field dependence of the Hall resistivity at different temperatures. In the whole temperature region, the Hall resistivity shows a non-linear field dependence, especially at $T = 1.4$ K and $T = 300$ K. The linear dependences at low fields are indicated by the dashed lines in Fig. 4(c). In the two-carrier model, the Hall resistivity $\rho_H$ is expressed as follows:

$$\rho_H = -\frac{\alpha}{\alpha^2 + \beta^2}, \tag{6}$$

where $\alpha$ and $\beta$ are:

$$\alpha = \frac{e_1 \sigma_1 \omega_{c1} \tau_1}{1 + \omega_{c1}^2 \tau_1^2} + \frac{e_2 \sigma_2 \omega_{c2} \tau_2}{1 + \omega_{c2}^2 \tau_2^2}, \tag{7}$$

$$\beta = \frac{\sigma_1}{1 + \omega_{c1}^2 \tau_1^2} + \frac{\sigma_2}{1 + \omega_{c2}^2 \tau_2^2}. \tag{8}$$

Here, $\sigma_i, \omega_{ci}$, and $\tau_i$ are the conductivity, cyclotron frequency, and the scattering time of band $i$. Further, $e_i = 1$ for electron carriers, and $e_i = -1$ for hole carriers. The conductivity $\sigma_i$ can be calculated as $\sigma_i = n_i e \mu_i$, where $n_i$ and $\mu_i$ are the carrier concentration and mobility of band $i$ respectively. The fitting of the Hall resistivity by Eq. (6) shows that there are two hole carriers at $T = 300$ K and one electron and one hole carrier at $T = 1.4$ K (see SM [27]). We attribute non-linearity of the field dependence of the Hall resistivity $\rho_H(B)$ at high temperatures to the coexistence of two hole pockets, i.e., the $L$ pocket and $\Sigma$ pocket, and that at low temperatures to the concave curvature of the Fermi surface [36]. The concave Fermi surface is consistent with our results of SdH oscillations. It can theoretically be shown that the concave curvature of the hole pocket leads to negative (not positive) Hall resistance [35]. Harrison et al.

explicitly showed that the concave curvature of the Fermi surface strongly changes the temperature dependence of the Hall resistivity [36]. As can be seen from Fig. 4(c), the temperature dependence of the Hall resistivity shows a different behavior at $T < 100$ K from that at $100 \text{ K} \leq T \leq 300$ K (the magnified view is shown in SM [27]). We attribute the slight temperature dependence of $R_H$ at $T < 100$ K to the concave curvature of the Fermi surface, and that at $100 \text{ K} \leq T \leq 300 \text{ K}$ to the carrier redistribution from the $L$ band to the $\Sigma$ band.

It is also theoretically shown that the actual carrier concentration of materials with concave Fermi surfaces can be correctly calculated from the saturation value of the Hall coefficient at the high-field limit: $R_{H,high\ B} = -1/(n_{eff}e)$, where $n_{eff} = n_e - n_h$ and $n_e$ and $n_h$ are the carrier concentration of electron/hole pockets, respectively [37]. The carrier concentration of Sr/Na codoped PbTe estimated from the Hall coefficient $R_H$ at low fields is $n_{Hall,low\ B} = 2.3 \times 10^{20}$ cm$^{-3}$ and $n_{Hall,high\ B} = 1.6 \times 10^{20}$ cm$^{-3}$ at high fields (see SM [27]), but the value of $R_H$ does not saturate up to 14.5 T. We can also estimate the carrier concentration $n$ from the Luttinger volume by roughly approximating the Fermi surface as ellipsoids with $K = 15$ and the carrier concentration is calculated as: $n_{Luttinger} = 0.9 \times 10^{20}$ cm$^{-3}$, which is closer to Na concentration determined by EPMA $[Na] = 0.7 \times 10^{20}$cm$^{-3}$. In short, because of the concave curvature of the Fermi surface in Sr/Na codoped PbTe, its carrier concentration is overestimated at low temperatures in the conventional analyses of the weak field Hall measurements, as suggested by earlier studies on heavily doped PbTe [39] [50]. The actual carrier concentration is difficult to determine, but it should be in the range of $0.9 \times 10^{20}$ cm$^{-3} < n < 1.6 \times 10^{20}$ cm$^{-3}$.

We consider that the concave Fermi surface will also affect the absolute value of the Seebeck coefficient. From the classical transport theory [37], the Seebeck coefficient is calculated as follows:

$$S = -\frac{\pi^2}{3}\frac{k_B^2 T}{e}\frac{\sigma'}{\sigma}, \quad (9)$$

where the energy derivative of the conductivity is calculated as follows:

$$\boldsymbol{\sigma}' = \frac{\tau'}{\tau}\boldsymbol{\sigma} + \frac{e^2\tau}{4\pi^3}\int d\boldsymbol{k}\delta(E_F - E(\boldsymbol{k}))\boldsymbol{M}^{-1}(\boldsymbol{k}). \quad (10)$$

Here, $E(\boldsymbol{k})$ and $\boldsymbol{M}(\boldsymbol{k})$ are the energy and the effective mass tensor at $\boldsymbol{k}$. When the energy dependence of the relaxation time $\tau$ is neglected, the Seebeck coefficient is proportional to the average of the inverse effective mass tensor $\boldsymbol{M}^{-1}$ on the Fermi surface. Therefore, the Fermi surface without a concave curvature is preferable to obtain a large Seebeck coefficient $S$ because contributions from different parts of the Fermi surface do not cancel out in the integral of Eq. (10). Since PbTe has two hole pockets ($L$ pocket and $\Sigma$ pocket) at high temperatures, the curvature of the $L$ pocket becomes more convex compared to the one-pocket case ($L$ pocket) with the same carrier concentration. We assume that the carrier redistribution by band convergence in PbTe makes its pockets more convex and this effect also contributes to the increased Seebeck coefficient $S$.

Next, we estimate the value of the energy separation between the Fermi energy and the edge of the $\Sigma$ band ($\Delta = \Delta E_{L\Sigma} - E_F$) from the temperature dependence of the Hall coefficient $R_H$. In this analysis, we extended the procedure of Allgaier [31] [38] (see SM [27]). When the Fermi energy crosses the first valence band and the edge of the second valence band is below $E_F$, the carrier distribution to the second valence band can be considered as an activation process. Allgaier et al. showed that the following relationship holds at weak fields when the conductivity of the second valence band is much smaller than that of the first valence band [31].

$$\frac{R_H - R_{H,L}}{R_{H,L}} = (1-b)^2 P e^{-\frac{\Delta}{k_B T}}, \qquad (11)$$

where $R_H$ and $R_{H,L}$ represent the Hall coefficient in the low-field limit and its saturation value at low temperatures, respectively. The other parameters are defined as $b = \mu_2/\mu_1$, $P = N_2 m_2^{\frac{3}{2}}/(N_1 m_1^{\frac{3}{2}})$, where $N_i$ and $m_i$ represents the number of the degenerate valleys and effective mass of the band $i$, respectively. In our analyses, we set $R_H$ as $R_H$ at $0\,\text{T} < B < 3\,\text{T}$ and set $R_{H,L}$ as $R_H$ at $T = 100$ K in order to eliminate the temperature dependence of $R_H$ due to the concave Fermi surface. Since Eq. (11) is rewritten as:

$$\ln\left(\frac{R_H - R_{H,L}}{R_{H,L}}\right) = -\frac{\Delta}{k_B T} + \ln(1-b)^2 P, \qquad (12)$$

one can obtain the energy separation $\Delta = \Delta E_{L\Sigma} - E_F$ from the Allgaier plot, that is, $\ln\left(\frac{R_H - R_{H,L}}{R_{H,L}}\right)$ vs $1/T$ plot. If there is a temperature dependence of $\Delta E_{L\Sigma}$, which is well approximated as $\Delta E_{L\Sigma} = \Delta E_{L\Sigma}|_{T=0} + A k_B T$ in PbTe [32] [40], we obtain the value of $\Delta$ at $T = 0$ K ($\Delta = \Delta_0$) in the plot. Although this procedure was originally applied to samples with low carrier concentrations [31], our numerical calculations in SM [27] show that this procedure can provide an accurate value of $\Delta_0$ even in samples with high carrier concentrations if the Fermi energy does not cross the second valence band. Details are discussed in SM [27]. Figure 4(d) shows the Allgaier plots for three different samples of the single crystals of Sr/Na codoped PbTe. Interestingly, the sample dependence of $\Delta_0$ is insignificant, indicating that Sr and Na are doped up to the solubility limit in our samples. The value $\Delta_0 \cong 50$ meV agrees well with the value obtained from the $dI/dV$ curve $\Delta_0 \cong 50$ meV in Fig. 2(b). This method can also be applied to polycrystalline samples, and hence we believe that it is beneficial to the field of TE research, where most studies are performed with polycrystals.

Finally, in order to study the temperature dependence of the carrier distribution quantitatively, we conduct the following analyses. On the basis of the two-carrier model, the weak-field limit of the Hall resistivity for two hole pockets is expressed as follows:

$$\rho_H = \frac{B}{e}\frac{n_1\mu_1^2 + n_2\mu_2^2}{(n_1\mu_1 + n_2\mu_2)^2} = \frac{B}{e}\frac{1 + \frac{n_2}{n_1}\frac{\mu_2^2}{\mu_1^2}}{n_1\left(1 + \frac{n_2\mu_2}{n_1\mu_1}\right)^2}. \tag{13}$$

In Sr/Na codoped PbTe, the first valence band ($L$ band, band 1) is considered to have much higher mobility than the second valence band ($\Sigma$ band, band 2) [31]. Then, the weak-field limit of the Hall resistivity in p-type PbTe can be well approximated as:

$$\rho_H = \frac{B}{en_1}. \tag{14}$$

Therefore, the carrier distribution in the $L$ pocket can be determined experimentally. Here, we assume that the total carrier concentration remains constant because the band gap ($E_G = 300$ meV) is larger than $3k_BT$ and activation of carriers to the conduction band is negligible. The result of the analyses is shown in Fig. 4(e), where we used $\rho_H$ at $0 < B < 3$ T (weak-field limit) and assumed $n_1|_{T=100\text{ K}} = n_{total}$ to eliminate the temperature dependence due to the concave Fermi surface. Here, we estimate that approximately 50% of the total carriers are distributed in the $\Sigma$ pocket at $T = 300$ K. This result can also be corroborated by the following theoretical analysis.

As now we have obtained the band parameters of Sr/Na codoped PbTe ($E_F = 100$ meV, $\Delta E_{L\Sigma} = 150$ meV, $E_G = 300$ meV). We can now estimate the carrier distribution between the $L$ band and the $\Sigma$ band using the Fermi distribution function. Based on classical semiconductor physics [38], the carriers are distributed by the following equations:

$$n_L \propto N_L \int_{E_L}^{\infty} \frac{4\pi \left(\frac{2m_L}{h^2}\right)^{\frac{3}{2}} (E-E_L)^{\frac{1}{2}}}{\exp\left(\frac{E-E_F}{k_B T}\right)+1} dE, \tag{15}$$

$$n_\Sigma \propto N_\Sigma \int_{E_\Sigma}^{\infty} \frac{4\pi \left(\frac{2m_\Sigma}{h^2}\right)^{\frac{3}{2}} (E-E_\Sigma)^{\frac{1}{2}}}{\exp\left(\frac{E-E_F}{k_B T}\right)+1} dE, \tag{16}$$

where $E_L$ and $E_\Sigma$ are the energies of the band edge of the $L$, $\Sigma$ band. In the calculation, the temperature dependence of $\Delta E_{L\Sigma}$ was set as $d\Delta E_{L\Sigma}/dT = 0.2$ meV/K, following the earlier studies [32] [39]. Details are described in SM [27]. This analysis can also be successfully utilized to calculate $T_{MAX}$, where $R_H(T)$ maximizes (see SM [27]). In the inset of Fig. 4(e), we plot $n_L/n_{total}$, which was calculated as a function of temperature. The calculated curve is in good agreement with the analyses of the experimental data. We note that the carrier concentration of the $L$ pocket, i.e., $n_L$ in Fig. 4(e) is overestimated (the carrier concentration of the $\Sigma$ band is underestimated) by the approximation $n_2\mu_2/n_1\mu_1 \cong 0$ in Eq. (13). In any case, almost all carriers are distributed in the $L$ band at low temperatures ($T < 100$ K). In contrast, about 50% of the total carriers are distributed in the $\Sigma$ band at $T = 300$ K because the $\Sigma$ band moves closer to $E_F$ at high temperatures [32]. The temperature dependence of the carrier distribution here also agrees well with the AMR results in Fig. 4(a). Therefore, we assume that the crossover behavior of the AMR at high temperatures reflects the redistribution of the carriers from the $L$ pocket to the $\Sigma$ pocket.

This study employed two distinct experimental techniques: $dI/dV$ measurements by STM and AMR measurements. Since the $dI/dV$ measurement can determine both the unoccupied and occupied band structure, we believe that it can be utilized to predict the physical properties of TE materials and it will provide a bird's eye view of the doping strategy of TE materials. In particular, when the energy dependence of the DOS is revealed, the optimal Fermi energy $E_F$ to maximize the Seebeck coefficient $S$ can be determined by the Boltzmann transport equation [49] if the energy dependence of the mobility is negligible. In addition, the doping effect on the band structure, such as the change of the band offset and band gap, can also be quantitatively determined as is shown in this study. Because many TE materials are easy to cleave, STM measurements on them are easy to perform. We believe that our methodology of $dI/dV$ measurements by STM can be applied to many other TE materials. We assume that the AMR can be applied to study the band structure of TE materials at high temperatures, which cannot be accessible by other experimental techniques such as STM and PES. Since the previous theoretical model [59] of the AMR does not include the effect of temperature, we consider that a more realistic model is needed to compare the experimental AMR data with theory at high temperatures. However, we can at least conclude that the AMR will be strongly modulated when the reconstruction of the Fermi surface occurs. We assume that the AMR should become a good indicator of temperature-induced band convergence.

## IV. CONCLUSION

STM and transport measurements have been performed to investigate the electronic structure and its temperature dependence in heavily Sr/Na codoped PbTe, which is recognized as one of the most promising TE materials. Our main findings are as follows: (i) Below $T = 4.5$ K, all carriers are distributed in the first valence band at the $L$ point ($L$ band), which forms tube-shaped Fermi surfaces with concave curvature. With Sr and Na doping, the dispersion of the $L$ band changes, and the band gap increases from $E_G = 200$ meV to $300$ meV. (ii) At $T = 4.5$ K, the Fermi energy is located $\sim 100$ meV below the edge of the $L$ band for Sr/Na codoped PbTe. The second valence band at the $\Sigma$ point ($\Sigma$ band) is lower than the $L$ band by $\Delta E_{L\Sigma} \cong 150$ meV, which is significantly smaller than that of pristine PbTe ($\Delta E_{L\Sigma} \cong 200$ meV). The decrease of $\Delta E_{L\Sigma}$, leading to band convergence, provides a desirable condition for TE materials. (iii) As the temperature increases, the carrier distribution to the $\Sigma$ band starts at $T \cong 100$ K, and we estimate that about 50% of the total carriers are distributed in the $\Sigma$ band at $T = 300$ K. Our work demonstrates that STM and AMR measurements are particularly powerful tools to experimentally investigate the correlation between the electronic structure and TE properties. We believe that they will provide a bird's eye view of the doping strategy towards high-efficiency TE materials.

## Acknowledgments

The authors are grateful to M. Nishio of Materials Analysis Station in NIMS for the EPMA measurements. YH acknowledges financial support by JSPS KAKENHI (Grant Nos. 21K20496 and 22K14467). SY by JSPS KAKENHI (Grant Nos. 20H05277, 21H01817). KS by the Murata Science Foundation. KE by JSPS KAKENHI (Grant No. 22H01765). SU by JSPS KAKENHI (22H01173). MANA was established by World Premier International Research Center Initiative (WPI), MEXT, Japan.

**Table**

Table I: Chemical composition (at.%) of Sr/Na codoped PbTe determined by EPMA

|  | Pb | Te | Sr | Na |
|---|---|---|---|---|
| Single crystal (Bridgman) | 49.2 | 49.8 | 0.6 | 0.4 |
| Polycrystal bulk matrix (Air-cooled) | 49.2 | 50.1 | 0.2 | 0.6 |

**Figure captions**

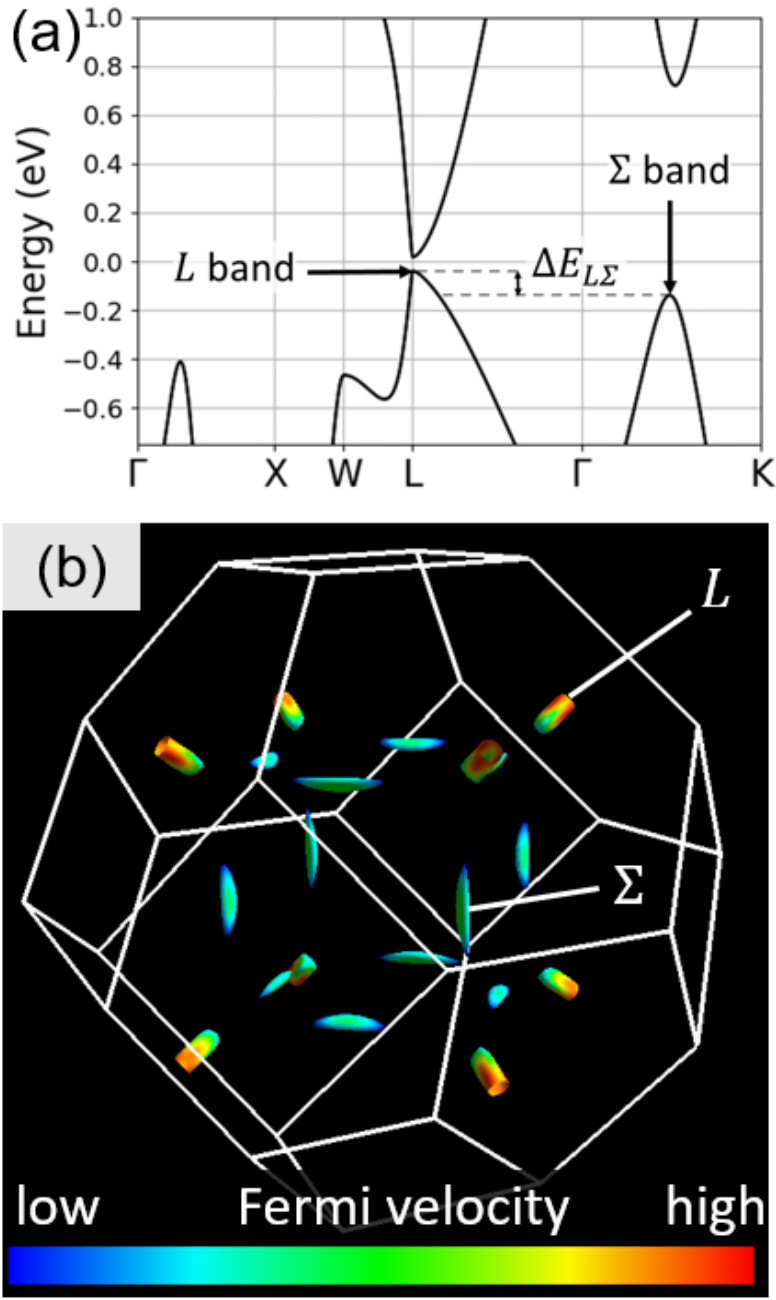

Fig. 1 (a): Calculated band structure of pristine PbTe. (b): The calculated Fermi surface of pristine PbTe at $E_F = 110$ meV below the edge of the $L$ valence band. Here, the white lines represent the first Brillouin zone.

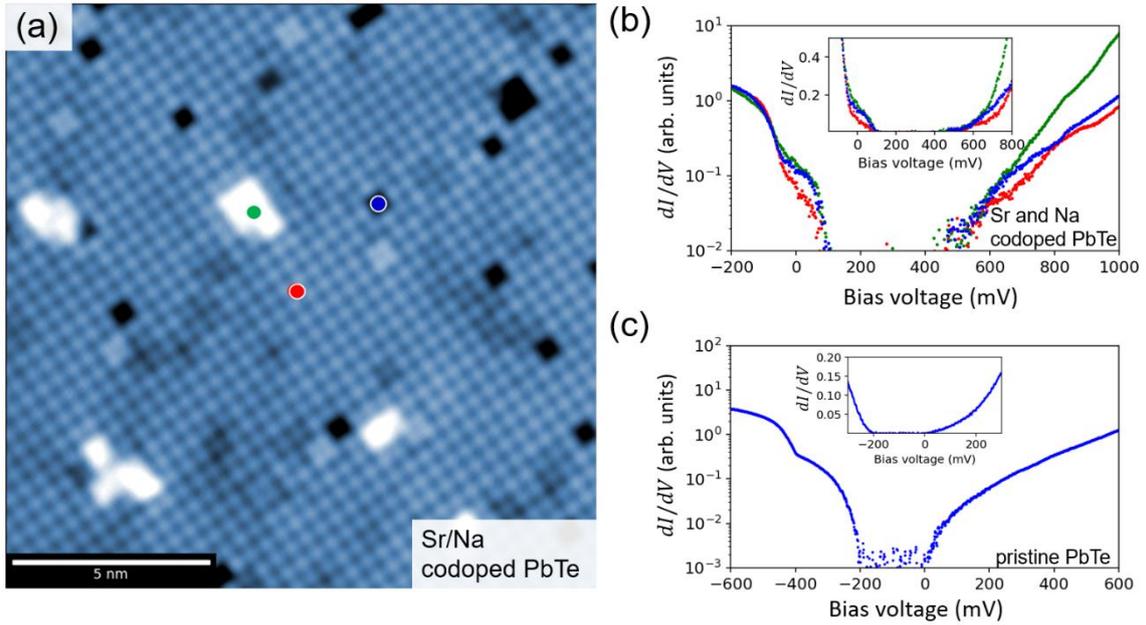

Fig. 2 (a): Scanning tunneling microscopy (STM) topography image on the (001) surface of single-crystalline Sr/Na codoped PbTe. The scale bar represents 5 nm. (b): $dI/dV$ curves measured at three points on the (001) surface of Sr/Na codoped PbTe (log-scale). The measurement positions are indicated in (a) by markers in the same colors as the $dI/dV$ curves. In the inset, the same curves are plotted on a linear scale. (c): The $dI/dV$ curve on the (001) surface of pristine and single-crystalline PbTe (log-scale). In the inset, the same curve is plotted on a linear scale. All measurements shown in Fig. 2 were performed at $T = 4.5$ K.

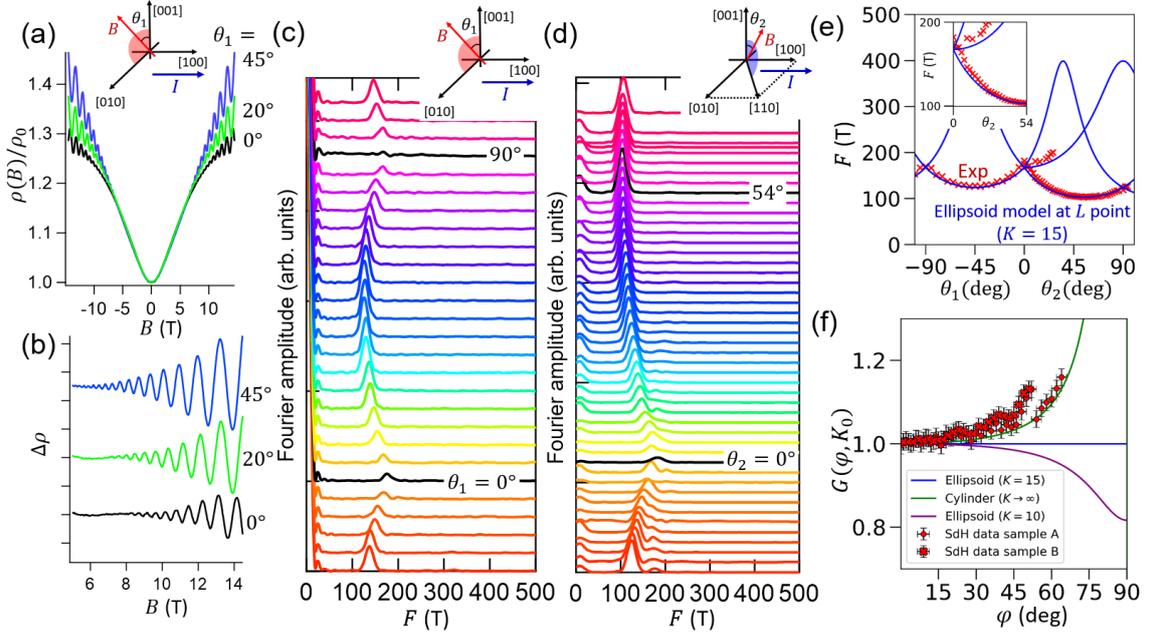

Fig. 3 (a): Magnetoresistance (MR) at different field angles. (b): The oscillatory components of the MR at different field angles. (c) and (d): Fourier transforms of Shubnikov–de Haas (SdH) oscillations at different angles. The measurements were performed at the field angle $\theta_1$ from [001] to [010] (angle step: 5°) in (c) and at $\theta_2$ from [001] to [110] (angle step: 2°) in (d). (e): Angular dependence of the SdH frequency in the experiments (c) and (d) together with one calculated from the ellipsoid model at $L$ point with $K = 15$. The inset shows an enlarged view. (f): Angular dependence of the shape function $G(\varphi, K_0)$. Here, we set $K_0 = 15$ and $\varphi$ is the angle between [111] and $\boldsymbol{B}$. The lines show calculated curves for the ellipsoidal pocket with $K = 15$ (blue), $K = 10$ (purple), and a cylindrical pocket (green). All the measurements in Fig. 3 were conducted in the single crystals of Sr/Na codoped PbTe at $T = 1.4$ K.

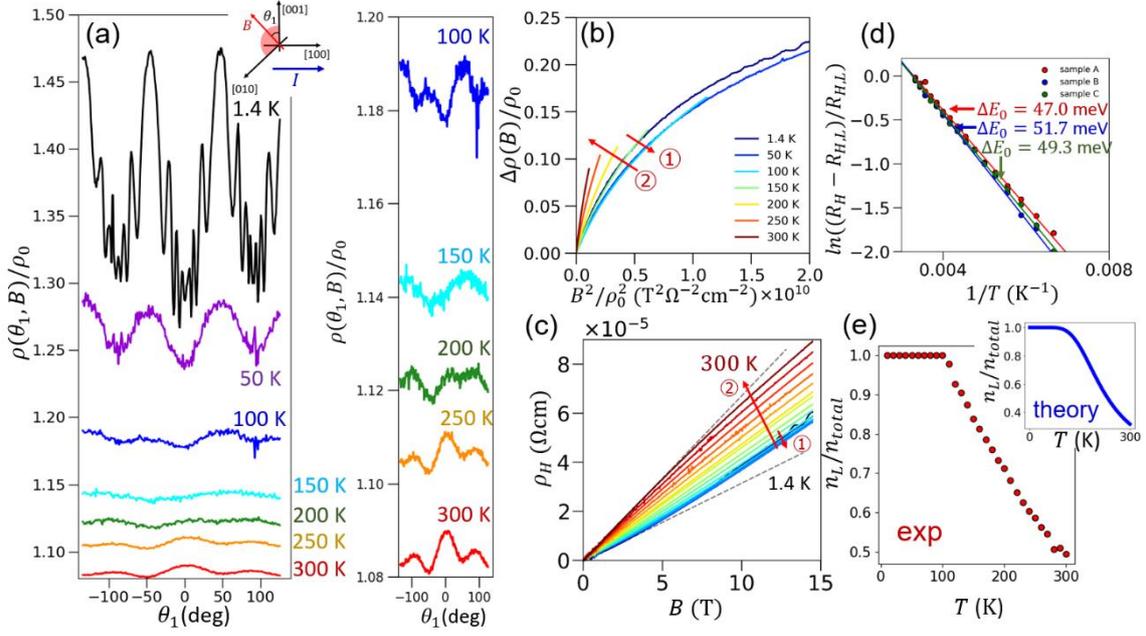

Fig. 4 (a): Angular dependent magnetoresistance (AMR) at different temperatures. Here, $B = 14.5$ T and the angle of the magnetic field is $\theta_1$. The definition of the angle $\theta_1$ is given in the inset. The right panel shows an enlarged view. (b): The Kohler plot at different temperatures for $\boldsymbol{B}//[001]$. (c): Magnetic field dependence of the Hall resistivity $\rho_H(B)$ at different temperatures. The dashed lines indicate the linear extrapolation of the low-field $\rho_H(B)$ at $T = 1.4$ and $300$ K. (d): The Allgaier plot for three different samples (see text). (e): Temperature dependence of the carrier distribution determined by Eq. (14). The inset shows the theoretical carrier distribution determined by Eqs. (15) and (16). All the measurements in Fig. 4 were conducted in the single crystals of Sr/Na codoped PbTe.

# Supplementary Materials:
# Experimental verification of band convergence in Sr and Na codoped PbTe

Yuya Hattori[1, 2, a)], Shunsuke Yoshizawa[3, b)], Keisuke Sagisaka[3], Yuki Tokumoto[2], Keiichi Edagawa[2], Takako Konoike[1], Shinya Uji[1] and Taichi Terashima[1]

[1]*Research Center for Materials Nanoarchitectonics (MANA), National Institute for Materials Science, 3-13 Sakura, Tsukuba, Ibaraki 305-0003, Japan*

[2]*Institute of Industrial Science, The University of Tokyo, Komaba, Meguro-ku, Tokyo 153-8505, Japan*

[3]*Center for Basic Research on Materials, National Institute for Materials Science, 1-2-1 Sengen, Tsukuba, Ibaraki 305-0047, Japan.*

Corresponding authors' E-mail: (a): HATTORI.Yuya@nims.go.jp (Yuya Hattori),

(b): YOSHIZAWA.Shunsuke@nims.go.jp (Shunsuke Yoshizawa)

Powder XRD

Figure S1 shows the powder XRD patterns for the single crystals of Sr/Na codoped PbTe (Bridgman method), the polycrystalline samples of Sr/Na codoped PbTe (air-cooled), and the single crystal of pristine PbTe (Bridgman method), together with simulated patterns of PbTe [1]. Although a small impurity peak is found at $2\theta \cong 31.3°$ in the air-cooled sample of Sr/Na codoped PbTe, we find no impurity peaks in the Bridgman sample of Sr/Na codoped and pristine PbTe, which we investigated in the main text. All the peaks of them can be indexed by the structure of PbTe [1] (space group: 225).

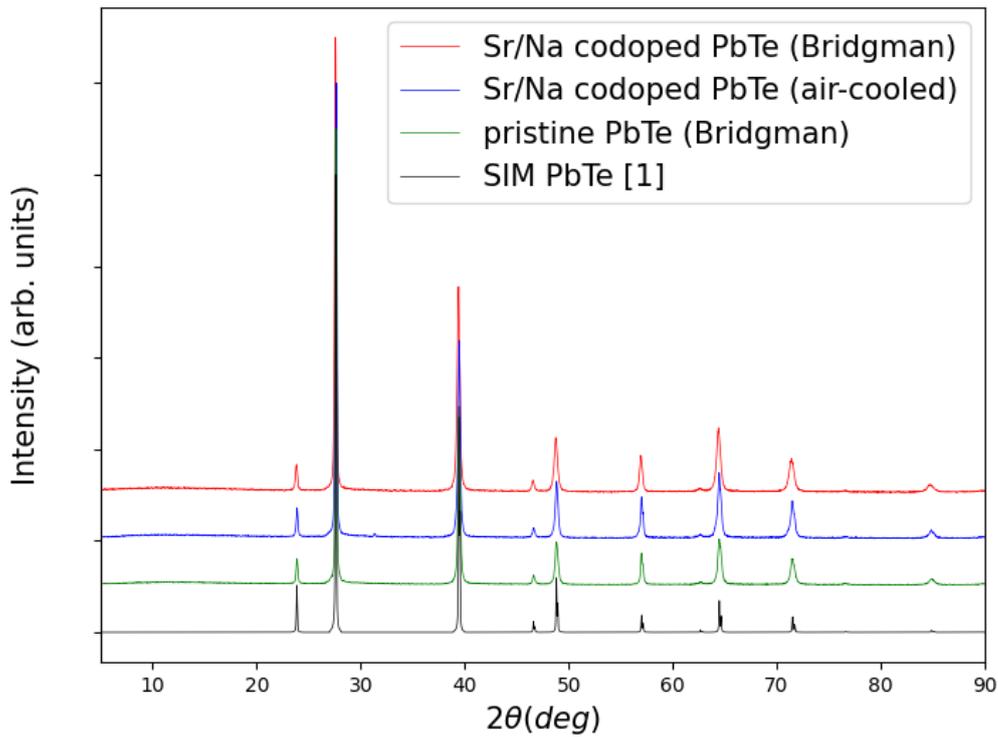

Fig. S1 The powder XRD patterns for the single crystals of Sr/Na codoped PbTe (Bridgman), polycrystalline sample of Sr/Na codoped PbTe (air-cooled), and the single crystals of pristine PbTe (Bridgman), together with the simulated pattern of PbTe.

Microstructure in polycrystalline samples of Sr and Na codoped PbTe (air-cooled)

Although no microstructure was observed in the single crystals of Sr/Na codoped PbTe (the Bridgman sample), the air-cooled samples showed specific microstructures. Figure S2 (a) shows the EPMA Sr elemental mapping in the air-cooled Sr and Na codoped PbTe samples.

There is a dendritic microstructure with high Sr concentration. In addition, we find a small Na-rich region inside the Sr-poor region (Fig. S2 (b)). The chemical compositions of the Sr-rich, Sr-poor (the bulk matrix), and Na-rich regions in Figs. S2 (a) and (b) are listed in Table SI. The chemical composition is averaged over five different probe spots in each region. The Sr concentration in the Sr-rich region ($\cong 10.5$ at.%) is surprisingly high and we infer from its composition that Sr atoms occupy the Pb site in PbTe. The Na concentration in the Na-rich region is approximately 4%. Although these high Sr and Na concentration regions occupy a small volume fraction in the crystal and do not affect the electrical conductivity $\sigma$ much, the electronic structure must be different from that of the bulk matrix. The boundary between the high concentration regions of Na and Sr and the bulk matrix should serve as a potential barrier for energy filtering [2] [3]. Such potentials are theoretically shown to increase the Seebeck coefficient $S$ if the Fermi energy is properly tuned.

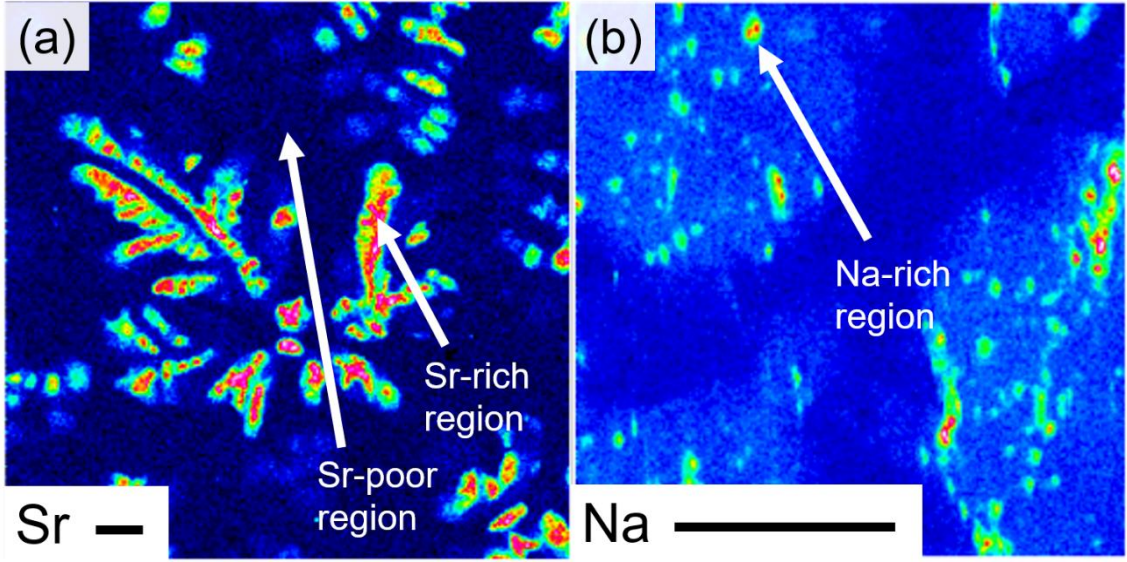

Fig. S2 EPMA (a) The Sr elemental mapping and (b) Na mapping in Sr/Na codoped PbTe (air-cooled). The scale bars in each figure represent $20 \mu m$.

|  | Pb | Te | Sr | Na |
|---|---|---|---|---|
| Sr-rich | 38.9 | 50.3 | 10.5 | 0.3 |
| Sr-poor (bulk matrix) | 49.2 | 50.1 | 0.2 | 0.6 |
| Na-rich | 48.4 | 46.5 | 1.2 | 3.9 |

Table SI: The chemical composition (at.%) of the air-cooled Sr and Na codoped PbTe.

Density of states of the Σ band in pristine and Sr and Na codoped PbTe

Figure S3 shows the $dI/dV$ curves for the single crystals of (a) pristine and (b) Sr and Na codoped PbTe. The spectra are quite well consistent with the earlier band calculation [4]. The values of the $dI/dV$ curves rapidly increase below $V \cong -400$ meV in pristine PbTe and $V \cong -50$ meV in Sr and Na codoped PbTe. The onsets correspond to the top of the Σ valence band. The sharp kink of the $dI/dV$ curves at the top of the Σ valence band is consistent with the earlier experimental studies of PbTe, which revealed the heavy mass of the Σ valence band [5] [6]. The energy dependence of the density of states is consistent with that expected in the parabolic band ($E \propto k^2$, $DOS \propto E^{0.5}$).

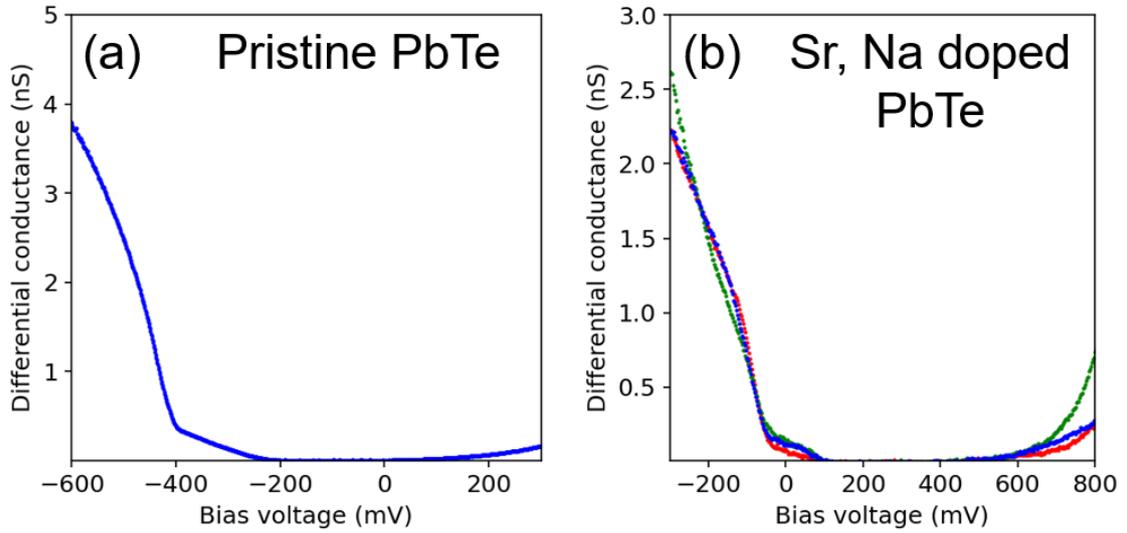

Fig. S3 $dI/dV$ curves for the single crystal of (a) pristine (b) Sr and Na codoped PbTe on a linear scale.

**Cyclotron effective mass in** Sr and Na codoped **PbTe single crystal**

In Figs. S4 (a) and (b), we present the temperature dependence of the SdH amplitudes of the $L$ pocket. Figure S4 (a) shows that along $\boldsymbol{B} \parallel [111]$ and Fig. S4 (b) along $\boldsymbol{B} \parallel [100]$. According to the Lifshitz-Kosevich (LK) formula, the temperature dependence of the amplitude of the SdH oscillations $A_{FFT}$ is expressed as:

$$A_{FFT} = \frac{A_0 \left( \frac{2\pi^2 k_B}{e\hbar} \frac{m^* T}{\bar{B}} \right)}{\sinh \left( \frac{2\pi^2 k_B}{e\hbar} \frac{m^* T}{\bar{B}} \right)}. \tag{S1}$$

The solid lines in Fig. S4 are the fits with the LK formula. The cyclotron effective mass along the [111] and [001] directions is $m^*_{111} = 0.11 m_e$ and $m^*_{100} = 0.21 m_e$ respectively, where $m_e$ is the free electron mass.

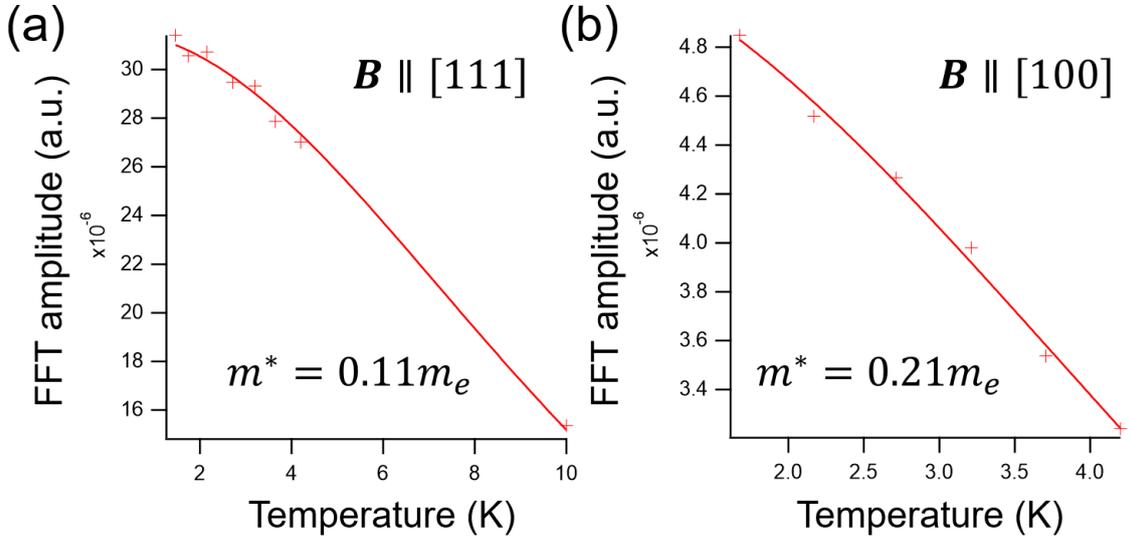

Fig. S4 The temperature dependence of the SdH amplitudes along (a) $\boldsymbol{B} \parallel [111]$ and (b) $\boldsymbol{B} \parallel [100]$ in the single crystal of Sr and Na codoped PbTe.

**The effect of the longitudinal magnetoresistance to the AMR in PbTe**

In their pioneering work on pristine PbTe, Sitter et al. investigated the angular dependent magnetoresistance (AMR) at different temperatures [7]. Their experimental setup was different from ours in the main text: the current direction was along $\bm{I} \parallel [110]$ and the magnetic field direction was tilted from $\bm{B} \parallel [001]$ to $\bm{B} \parallel [110]$ (for details, see [7]). The magnetoresistance (MR) then includes both the components of the longitudinal and transverse MR. The AMR of this setup had the periodicity of $\Delta\theta = 180°$ (a two-fold symmetry) [7]. We consider that a small component of a two-fold symmetry in Fig. 4 (a) in the main text originates from the component of the longitudinal MR, which appears due to a slight misalignment in the experiment. In PbTe, the longitudinal MR is known to be much larger than the transverse MR [7] [8]. In support of this notion, the AMR measured at $B = 14.5$ T can be well fitted by sinusoidal waves with a period of $\Delta\theta = 90°$ (four-fold symmetry, the transverse MR) and $\Delta\theta = 180°$ (two fold symmetry, the longitudinal MR) as shown in Fig. S5. In the fitting, the experimental AMR was fitted by the following function.

$$\frac{\rho(\theta, B)}{\rho_0} = a_0 + a_2 \sin(2\theta + \phi_2) + a_4 \sin(4\theta + \phi_4). \tag{S2}$$

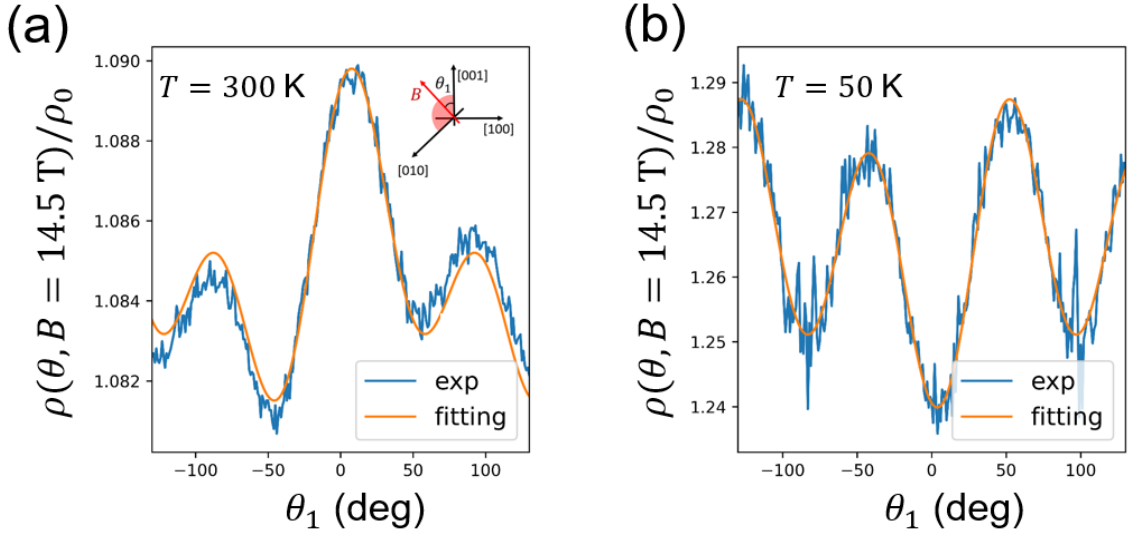

Fig. S5 The AMR of the single crystal of Sr and Na codoped PbTe, together with the fitted curves at (a) $T = 300$ K and (b) $T = 50$ K. The magnetic field was $B = 14.5$ T both in (a) and (b).

### The Hall resistivity at high fields

Figure S6 shows the field dependence of the Hall resistivity $\rho_H$ at different temperatures, and it is the enlarged view of the Fig. 4(c) in the main text. At the higher temperatures (100 K ≤ $T$ ≤ 300 K), the slope of $\rho_H$ increases with increasing temperature. This behavior is due to the carrier redistribution from the $L$ band to the $\Sigma$ band. At the lower temperatures ($T$ < 100 K), in contrast, the behavior is different from that at the higher temperatures as shown by the arrows in Fig. S6. We attribute it to the concave Fermi surface. When the Fermi surface consists of concave curvature, its Hall resistivity $\rho_H$ is known to show specific temperature dependence due to the partial movement around cyclotron orbits on the Fermi surface [9] [10].

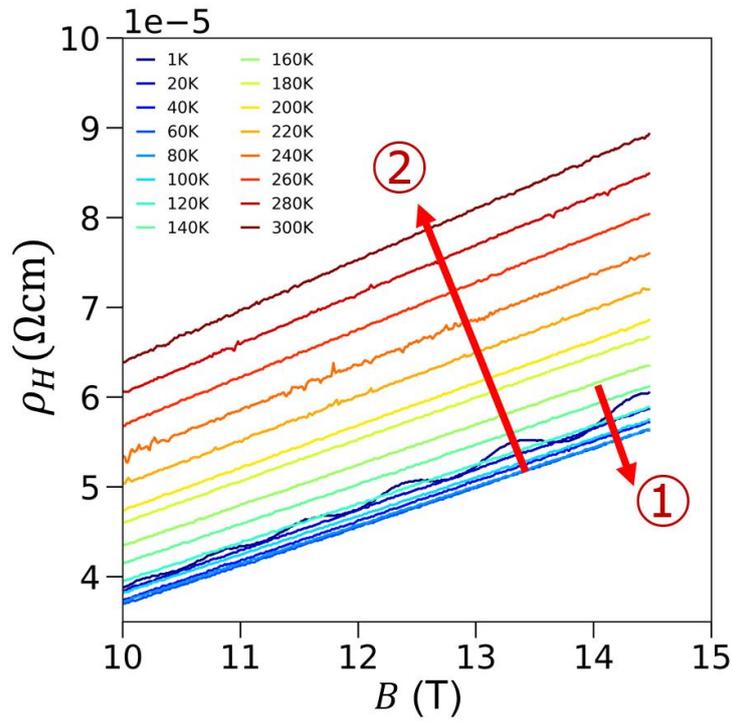

Fig. S6 The field dependence of $\rho_H$ at different temperatures. The sample is the single crystal of Sr and Na codoped PbTe.

## The saturation of the Hall coefficient at high fields

It can be shown theoretically that the high-field limit of the Hall coefficient $R_H$ is directly related to the total carrier concentration [11]. If there is only hole pocket, the relation is: $\lim_{B\to\infty} R_H = 1/(n_h e)$, where $n_h$ is the hole carrier concentration. The field dependence of the Hall coefficient in the single crystal of Sr and Na codoped PbTe at $T = 1.4$ K is shown in Fig. S7. The oscillating components are from SdH oscillations. We calculate $n_{Hall, highB} = 1.6 \times 10^{20}$ cm$^{-3}$ from the $R_H$ value at $B = 14.5$ T, but the actual value of the total carrier concentration must be slightly smaller than this value because $R_H$ does not saturate at $B = 14.5$ T.

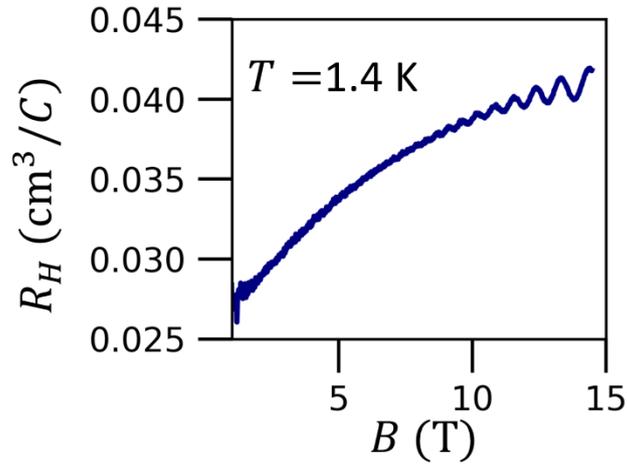

Fig. S7 The field dependence of $R_H$ in the single crystal of Sr and Na codoped PbTe. The temperature is $T = 1.4$ K.

## The two-carrier model analyses of the Hall resistivity

Figure S8 shows the fitting of the Hall resistivity $\rho_H$ by the two-carrier model for single-crystalline Sr and Na doped PbTe, where the fitting function was Eq. (6) in the main text. The Hall resistivity was fitted by one hole carrier and one electron carrier at $T = 1.4$ K and two hole carriers at $T = 300$ K. The fitting parameters for each temperature are: $n_{h1} = 1.5 \times 10^{20}$ cm$^{-3}$, $\mu_{h1} = 16000$ cm$^2$V$^{-1}$s$^{-1}$, $n_{e2} = 4.5 \times 10^{19}$ cm$^{-3}$, $\mu_{e2} = 2000$ cm$^2$V$^{-1}$s$^{-1}$ at $T = 1.4$ K and $n_{h1} = 5.0 \times 10^{19}$ cm$^{-3}$, $\mu_{h1} = 1000$ cm$^2$V$^{-1}$s$^{-1}$, $n_{h2} = 1.5 \times 10^{20}$ cm$^{-3}$, $\mu_{h2} = 75$ cm$^2$V$^{-1}$s$^{-1}$ at $T = 300$ K.

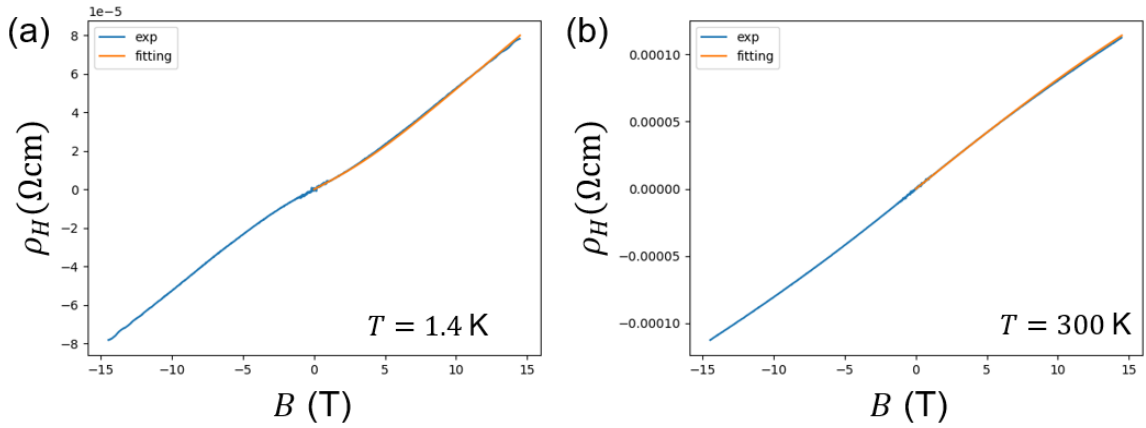

Fig. S8 The fitting of the Hall resistivity by the two-carrier model for single-crystalline Sr and Na codoped PbTe at (a) $T = 1.4$ K and (b) $T = 300$ K.

## The temperature dependence of Hall coefficient $R_H$ and the Allgaier plot

Here, we show numerically when the analysis of Allgaier [5] [12] can be applied to estimate $\Delta = \Delta E_{L\Sigma} - E_F$, which is the energy separation between the Fermi energy and the top of the second valence band in PbTe. The carrier concentration at finite temperature $T$ can be calculated from the classical semiconductor physics [12]:

$$n_L(T) \propto N_L \int_{E_L}^{\infty} \frac{4\pi \left(\frac{2m_L}{h^2}\right)^{\frac{3}{2}} (E - E_L)^{\frac{1}{2}}}{\exp\left(\frac{E - E_F}{k_B T}\right) + 1} dE = f_L(T), \quad (S3)$$

$$n_\Sigma(T) \propto N_\Sigma \int_{E_\Sigma}^{\infty} \frac{4\pi \left(\frac{2m_\Sigma}{h^2}\right)^{\frac{3}{2}} (E - E_\Sigma)^{\frac{1}{2}}}{\exp\left(\frac{E - E_F}{k_B T}\right) + 1} dE = f_\Sigma(T), \quad (S4)$$

where $N_L$ and $N_\Sigma$ are the valley degeneracy and $N_L = 4, N_\Sigma = 12$ in PbTe. The parameters $m_i$ and $E_i$ represent the effective mass and the energy of band edge of band $i$. We assume that the total carrier concentration $n_0 = n_L + n_\Sigma$ does not show the temperature dependence and takes a constant value because there is almost no activation process between the valence band and conduction band due to the large band gap ($E_G \cong 300$ meV at $T = 4.2$ K). Then,

$$n_L(T) = n_0 \times \frac{f_L(T)}{f_L(T) + f_\Sigma(T)}, \tag{S5}$$

$$n_\Sigma(T) = n_0 \times \frac{f_\Sigma(T)}{f_L(T) + f_\Sigma(T)}. \tag{S6}$$

The Hall coefficient $R_H$ at the low-field limit is calculated based on the two carrier model as:

$$R_H = \frac{1}{e} \frac{n_L \mu_L^2 + n_\Sigma \mu_\Sigma^2}{(n_L \mu_L + n_\Sigma \mu_\Sigma)^2}, \tag{S7}$$

where $n_i$ and $\mu_i$ represent the carrier concentration and mobility of band $i$. At low temperatures, all carriers are located at the $L$ point and $R_H$ saturates to the specific value: $R_{H,L}$ [5]. Using Eqs. (S3-S7), we plot $\ln\left(\frac{R_H - R_{H,L}}{R_{H,L}}\right)$ vs $1/T$ (dubbed Allgaier plot) with different values of $\Delta_0 = (\Delta E_{L\Sigma} - E_F)|_{T=0K}$ in Figs. S8 (a-d). In the calculation, the parameters are set as follows: $\Delta E_{\Sigma L} = 150 - 0.2T$ meV [13], $\frac{m_\Sigma}{m_L} = 2.68$ [14] and $\frac{\mu_\Sigma}{\mu_L} = 0.2$ [7]. From Fig. S9, we can see that when $\Delta_0 = (\Delta E_{L\Sigma} - E_F)|_{T=0K} > 0$, namely Fermi level does not cross the second valence band, the energy separation $\Delta$ can be estimated with good accuracy by this procedure (Figs. S9 (a-c)). In contrast, when the Fermi energy crosses the second valence band ($\Delta_0 < 0$), the plot does not show the linear relation as shown in Fig. S9 (d). This is further evidence that the Fermi energy in the single crystal of Sr and Na codoped PbTe does not cross the second valence band at low temperatures. In the analyses of the experimental data, we set $R_{H,L}$ as $R_H$ at $T = 100$ K because we find a small temperature dependence of $R_H$ at $T < 100$ K due to the concave shape of the Fermi surface.

This procedure also allows us to study the temperature dependence of the Hall coefficient $R_H(T)$, which has been used qualitatively to investigate the energy separation $\Delta E_{L\Sigma}$ in the study of PbTe [15] [16]. We find that the temperature $T = T_{MAX}$, where Hall coefficient $R_H$ maximizes, is reduced if we decrease the value of band offset $\Delta E_{L\Sigma}$ (Fig. S9 (e)). Since the temperature at which $R_H$ maximizes is when the conductance of two valence bands becomes equal [5], the behavior seems reasonable. The value of $T_{MAX}$ becomes comparable to that reported in polycrystalline Sr and Na codoped PbTe [15] when we assume $\Delta E_{L\Sigma}|_{T=0K} = 150$ meV and $E_F = 100$ meV as shown in Fig. S9 (e). It also confirms the validity of the electronic structure obtained in this study. Since $R_H(T)$ in our study did not reach the saturation value below $T = 300$ K, the conductivity of the $L$ band seems still higher than that of the $\Sigma$ band at $T = 300$ K.

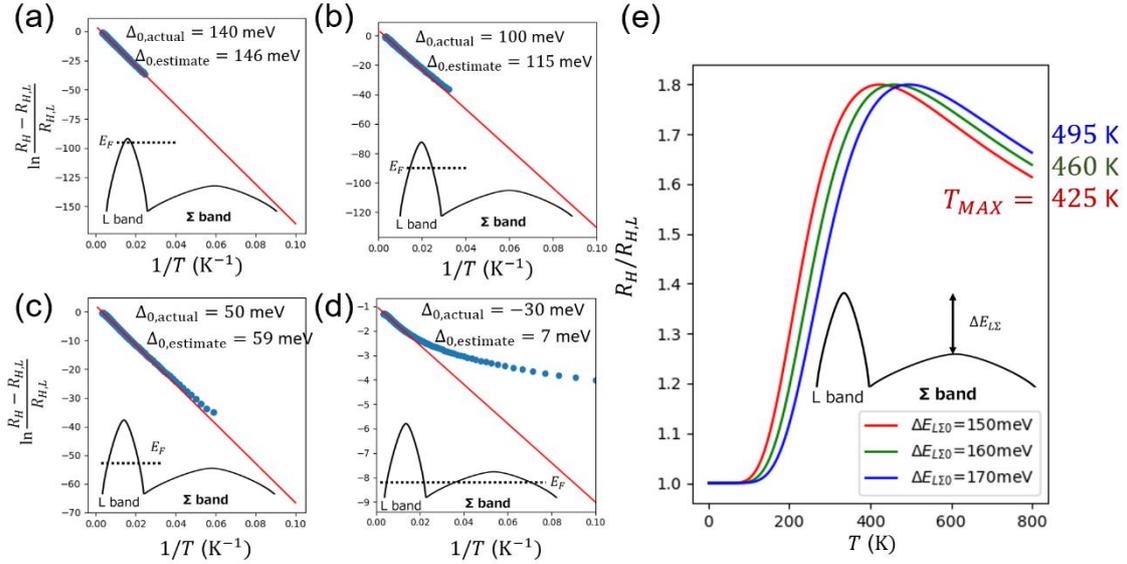

Fig. S9 (a-d) The Allgaier plots for the simulated data with different Fermi energies (circles); (a)$\Delta_0 = 140$ meV, (b)$\Delta_0 = 100$ meV, (c)$\Delta_0 = 50$ meV, and (d)$\Delta_0 = -30$ meV. The red lines are linear fits to the plots. (e) The calculated temperature dependence of Hall coefficient $R_H(T)$, which takes its maximum at $T = T_{MAX}$.